\documentclass[twocolumn,showpacs]{revtex4}
\usepackage{epsf,amsfonts,latexsym,amsmath,latexsym,amssymb}

\newtheorem{definition}{Definition}
\newtheorem{theorem}{Theorem}

\def\C{{\mathbb{C}}}

\def\N{{\mathbb{N}}}
\newcommand{\F}[0]{{\mathbb{F}}}

\newcommand{\cC}[0]{{\mathcal C}}
\newcommand{\cU}[0]{{\mathcal U}}

\newcommand{\cD}[0]{{\mathcal D}}

\newcommand{\cH}{{\cal H}}

\newcommand{\cQ}{{\cal Q}}

\newcommand{\DFT}[0]{{\rm DFT}}

\newcommand{\onemat}[0]{{\mathbf 1}}
\newcommand{\zeromat}[0]{{\mathbf 0}}

\newcommand{\proof}[1]{{\bf Proof.} #1 \hfill $\Box$\vspace{0.5cm}}

\newcommand{\ket}[1]{|#1\rangle}
\newcommand{\bra}[1]{\langle #1|}

\begin{document}

\title{Efficient decoupling schemes with bounded
controls\\ based on ``Eulerian'' orthogonal arrays}

\author{Pawel Wocjan}
\email{wocjan@cs.caltech.edu}
\affiliation{Institute for Quantum Information\\ 
California Institute of Technology\\ 
Pasadena, California 91722}

\begin{abstract}
The task of decoupling, i.e., removing unwanted interactions in a
system Hamiltonian and/or couplings with an environment (decoherence),
plays an important role in controlling quantum systems.  There are
many efficient decoupling schemes based on combinatorial concepts like
orthogonal arrays, difference schemes and Hadamard matrices. So far
these (combinatorial) decoupling schemes have relied on the ability to
effect sequences of instantaneous, arbitrarily strong control
Hamiltonians (bang-bang controls). To overcome the shortcomings of
bang-bang control {\sc Viola and Knill} proposed a method called
``Eulerian decoupling'' that allows the use of bounded-strength
controls for decoupling. However, their method was not directly
designed to take advantage of the composite structure of multipartite
quantum systems. In this paper we define a combinatorial structure
called an Eulerian orthogonal array. It merges the desirable
properties of orthogonal arrays and Eulerian cycles in Cayley graphs
(that are the basis of Eulerian decoupling). We show that this
structure gives rise to decoupling schemes with bounded-strength
control Hamiltonians that can be applied to composite quantum systems
with few body Hamiltonians and special couplings with the
environment. Furthermore, we show how to construct Eulerian orthogonal
arrays having good parameters in order to obtain efficient decoupling
schemes.
\end{abstract}

\pacs{{03.67.Lx, 03.65.Fd, 03.67.-a}}% PACS, the Physics and Astronomy
                             % Classification Scheme.

\maketitle

\section{Introduction}
An important task in quantum information theory consists in
selectively removing unwanted contributions of the system Hamiltonian
and/or switching off couplings of the quantum system to an
uncontrollable environment (the later being responsible for
decoherence). This task is usually called decoupling (see e.g.\
\cite{bangbang,violaDecoupling,violaControl,Zandari:00} and e.g.\
\cite{knill,stoll,leung,finite,reversal} for schemes using
combinatorial concepts). More generally, one is also interested in
effectively changing the system Hamiltonian in order to simulate some
desired Hamiltonian; this is usually referred to as simulating
Hamiltonians (see e.g.\
\cite{WJB:02,DNBT:2002,finite,BCLLLPV:2002}). In this paper we will
concentrate on designing efficient decoupling schemes.

Methods of dynamical decoupling and also simulating Hamiltonians
derive their basic physical intuition from coherent averaging
techniques in high-resolution nuclear magnetic resonance (NMR)
spectroscopy \cite{WHH:68,EBW:87}. A decoupling scheme is understood
as a control protocol which relies on the repeated application of
controls drawn from a finite set in order to change effectively the
natural time evolution to the desired evolution. Many efficient
decoupling schemes can be designed with the help of combinatorial
concepts like e.g.\ Hadamard matrices, difference schemes and
orthogonal arrays. The entries of these structures describe how to
choose the controls. The reason why it is possible to use these
combinatorial objects is the special structure of the system
Hamiltonians (pair-interactions or more generally few body
Hamiltonians). So far all these (combinatorial) schemes relied on the
ability to effect sequences of instantaneous, arbitrarily strong
controls Hamiltonians (bang-bang controls). Because of the requirement
of bang-bang controls such schemes are unrealistic in many
situations. To overcome the shortcomings of bang-bang decoupling {\sc
Viola and Knill} proposed a general method (called Eulerian
decoupling) for implementing decoupling with bounded controls (i.e.,
continuously modulated bounded-strength control Hamiltonians)
\cite{VK:02,Viola:04}. This method offers many advantageous over
bang-bang decoupling. However, their method was not directly designed
to make use of the special structure of few body Hamiltonians to
reduce the complexity of decoupling.

We show how to incorporate some of the above combinatorial methods
(that were used so far only in the bang-bang formulation) into the
method by {\sc Viola and Knill} in order to obtain efficient
decoupling schemes with bounded controls. Our schemes can be applied
to few body Hamiltonians and special couplings with an
environment. Our schemes rely on a combinatorial object (which we
call) an Eulerian orthogonal array. We show how to construct these
objects with good parameters from error correcting codes.

The paper is organized as follows. In section~2 we describe the
principles of dynamical decoupling. We describe briefly the so-called
first-order approximation that is the basis for all decoupling
schemes. In section~3 we recall how to construct decoupling schemes
with bang-bang and bounded strength controls. The method using bounded
strength control is {\sc Viola and Knill}'s Eulerian decoupling. Here
no special structure of the quantum system is assumed. Then in
section~4 we consider quantum systems consisting of coupled qudits. We
first recall in subsection~4.1 how to construct efficient decoupling
schemes with bang-bang controls with the help of orthogonal
arrays. Efficiency means that the number of necessary pulses grows
polynomially with the number of qudits. In subsection~4.2 we show how
to merge the concept of orthogonal arrays with the idea of Eulerian
decoupling in order to obtain efficient decoupling schemes with
bounded controls. Our method is based on a combinatorial structure
called Eulerian orthogonal array. In sections~5 we show how to
construct Eulerian orthogonal arrays with good parameters.

\section{Principles of dynamical decoupling}
A decoupling scheme is understood as a control protocol which relies
on the repeated application of controls drawn from a finite set in
order to change effectively the natural time evolution to the desired
evolution. We refer the reader to \cite{VK:02,Viola:04} for a more
detailed description. In the following we give a brief introduction
based on the above articles.

The joint evolution of the target system $S$ in interaction with the
environment $E$ is described by a total drift Hamiltonian of the form
\begin{equation}
H=H_S\otimes\onemat_E + \onemat_S \otimes H_E + H_{SE}\,,\,\,\,
H_{SE} = \sum_a S_a \otimes E_a\,,
\end{equation}
where $H_S$ and $H_E$ characterize the isolated dynamics of the system
and the environment, respectively, and the interaction term $H_{SE}$
is responsible for introducing unwanted decoherence effects and
dissipation effects in the reduced dynamics of $S$ alone. Without loss
of generality we will always choose the operators $S_a$ and $H_S$ to
be traceless.

The idea behind dynamical decoupling is to add a specially designed
controller, described by a time-dependent control Hamiltonian $H_c(t)$
acting on only the target system $S$, in such a way that the resulting
controlled dynamics is described by an effective Hamiltonian $H_{\rm
eff}$ which no longer contains any coupling terms between $S$ and $E$,
i.e.,
\begin{equation}
H_{\rm eff} = \tilde{H}_S\otimes\onemat_E + \onemat_S\otimes H_E\,,
\end{equation}
for an appropriate, possibly modified, system Hamiltonian
$\tilde{H}_S$. In this paper we will be interested in the case that
$\tilde{H}_S=\zeromat$.

Decoupling protocols are most conveniently constructed by directly
looking at the control propagator associated to $H_c(t)$ 
\begin{equation}
U_c(t)={\cal T}\exp\left\{ -i \int_0^t H_c(\tau) d\tau\right\}\,,
\end{equation}
where ${\cal T}$ denotes the time ordering.

Decoupling is based on the so-called {\em first order decoupling}. The
control actions are always cyclic, i.e., $U_c(t+T_c)=U_c(t)$ for some
cycle time $T_c$ and for all $t$.  The stroboscopic dynamics $U(t_M)$
with $t_M=M T_c$ and $M\in\mathbb{N}$ may be described by a propagator
\begin{equation}
U(t_M)=\exp(-i \bar{H} t_M)
\end{equation}
for a {\em time-independent} effective Hamiltonian $H_{\rm eff}$.
If, in addition, $T_c$ is sufficiently short, then the effective Hamiltonian
is accurately represented by the following lowest-order Hamiltonian 
\begin{equation}\label{eq:approx}
\bar{H}=\frac{1}{T_c}
\int_0^{T_c} U_c^\dagger(t) H U_c(t)\, d\tau\,.
\end{equation}
While higher-order terms can be systematically evaluated, the
approximation in (\ref{eq:approx}) becomes more and more exact as the
fast control limit $T_c\rightarrow 0$ is approached. First-order
decoupling is based on this approximation.

Having introduced the framework of decoupling we address the problem
of designing efficient decoupling schemes with first bang-bang and
then bounded-strength controls for general Hamiltonians and couplings
with the environment.

\section{Decoupling schemes}
In this section do not assume any special structure of the target
system $S$, the Hamiltonian $H_S$, and the coupling to the environment
given by $S_a$'s. We first discuss how to realize decoupling with
controls of unbounded strength (bang-bang) and then with bounded
control (Eulerian decoupling). The presentation is based on
\cite{VK:02}.

\subsection{Bang-bang control}
The time-average in (\ref{eq:approx}) can be expressed directly as an
average over a group in the following simple bang-bang decoupling
setting. Let $G$ be a discrete group of order $|G|$ acting on the
Hilbert space of the target system $\cH_S$ via a faithful, unitary,
projective representation,
\begin{equation}\label{eq:representation}
\rho : 
\left\{
\begin{array}{ccc} 
G & \rightarrow & {\cal U}({\cH_S}) \\
g & \mapsto     & U_g:=\rho(g)
\end{array}
\right.\,,
\end{equation}
where ${\cal U}(\cH_S)$ denotes the group of unitary matrices acting
on $\cH_S$. Let $\lambda\in\mathbb{N}$ and $\lambda>0$. 

A decoupling scheme using $U_g$'s as control operations is specified
by a sequence $(g_1,g_2,\ldots,g_N)$ with $N:=\lambda |G|$ and
entries from $G$. The number $N$ is called the length
of the decoupling scheme. The entries $g_j$'s specify the control
propagator $U_c(t)$ over each of the $\lambda |G|$ equally long
subintervals. A control cycle is defined by
\begin{equation}\label{eq:control}
U_c\left( (j-1) \Delta  + \tau\right) = U_{g_j}\,,\quad \tau\in [0,\Delta)\,,
\end{equation}
with $T_c=\lambda |G|\Delta$ for some $\Delta > 0$, and
$j=1,\ldots,N$.

If all group elements appear exactly $\lambda$ times in the list
$(g_1,\ldots,g_N)$ then the resulting control action corresponds
to extracting the $G$-invariant component of $X$. We have
\begin{equation*}
\frac{1}{N}\sum_{j=1}^N
U_{g_j}^\dagger X U_{g_j} =
\frac{\lambda}{N}\sum_{g\in G} U_g^\dagger X U_g = \Pi_G(X)\,,
\end{equation*}
where
\begin{equation}\label{eq:groupAverage}
\Pi_{G}(X) = \frac{1}{|G|} \sum_{g\in G} U_g^\dagger X U_g\,.
\end{equation}
Note that if the representation $\rho$ in (\ref{eq:representation}) is
irreducible than we have $\Pi_G(X)=tr(X)/d\,\, \onemat_d$ for all $X$,
where $d$ is the dimension of $\cH_S$.

An example for such an irreducible, unitary, projective representation
is given in the following. The discrete Fourier transform of length $d
\in \N$ is the unitary transformation defined by $\DFT_d :=
\frac{1}{\sqrt{d}} \sum_{k, \ell=0}^{d-1} \omega^{k \cdot \ell}
\ket{k}\bra{\ell}$, where $\omega$ denotes the primitive $d$-th root
of unity $e^{2 \pi i/d}$.  Next, define operators $S :=
\sum_{k=0}^{d-1} \ket{k}\bra{k+1}$, where the indices are reduced
modulo $d$, and $T := \DFT_d^{\dagger} \cdot S \cdot \DFT_d =
\sum_{k=0}^{d-1} \omega^k \ket{k}\bra{k}$. Then the map
\begin{equation}\label{eq:heisenberg}
\rho : \left\{
\begin{array}{ccc}
Z_d \times Z_d & \rightarrow & {\cal U}(d) \\
(a,b)          & \mapsto     & S^a T^b
\end{array}
\right.
\end{equation}
is an irreducible, unitary, projective representation. Note that for
$d=2$ one obtains $\onemat,\sigma_x,\sigma_y,\sigma_z$, where the
$\sigma$'s are the Pauli matrices.

Now it clear that we can remove the couplings with the environment and
switch off the natural time evolution of the quantum system by
performing the control operations according to (\ref{eq:control}) and
the representation in (\ref{eq:heisenberg}). This is because
$\Pi_G(S_a)=\zeromat$ for all $a$ and $\Pi_G(H_S)=\zeromat$.  However,
this method has the following disadvantage that makes it unrealistic
form in many situations. According to the rule in (\ref{eq:control})
the control propagator $U_c(t)$ jumps from $U_{g_{j-1}}$ to
$U_{g_j}=(U_{g_j} U_{g_{j-1}}^\dagger) U_{g_j}$ through the
application of an arbitrarily strong, instantaneous kick at the $j$th
endpoint $t_j = j \Delta$, realizing the bang-bang pulse $U_{s_j} =
U_{g_j} U_{g_{j-1}}^\dagger$ with $s_j:=g_j g_{j-1}^{-1}$ (equality is
understood here up to a phase factor). In the next section we describe
how to avoid such bang-bang controls.

\subsection{Eulerian decoupling}
As already mentioned in the introduction the requirements for
bang-bang control are highly unrealistic. {\sc Viola and Knill}
proposed a method called Eulerian decoupling \cite{VK:02} that avoids
the use of such bang-bang pulses. In Eulerian decoupling the control
propagator $U_c(t)$ is varied smoothly from $U_{g_{j-1}}$ to $U_{g_j}$
by a control action distributed along the whole $j$th subinterval.

Let $S$ be a generating set for $G$, i.e., any element of $G$ can be
written as a product of elements of $S$. The Cayley graph
$\Gamma(G,S)$ of $G$ with respect to $S$ is a directed graph whose
vertices are labeled by the group elements and whose edges are labeled
by the generators. More precisely, the vertex $g$ is joined to the
vertex $h$ if and only if $g h^{-1}=s$ for some $s\in S$, i.e., $g=s
h$.

It is assumed that we have the ability to physically implement the
generators $s\in S$, i.e., to implement the unitaries $U_s$ by the
application of some suitably chosen control Hamiltonians $h_s(t)$ over
$\Delta$:
\begin{equation}\label{eq:control1}
U_s = u_s(\Delta)
\end{equation}
where
\begin{equation}\label{eq:control2}
u_s(\delta) = 
{\cal T} \left\{ 
\exp\big( -i\int_0^\delta h_s(\tau) d\tau \big)
\right\}
\end{equation}
for $\delta\in [0,\Delta]$. The choice of the control Hamiltonians
$h_s(t)$ is not unique. This allows for additional flexibility for the
concrete implementation. Once a choice of the control Hamiltonians is
made, the control action is determined by assigning a cycle time and a
rule for switching the Hamiltonians $h_s(t)$ during the cycle
subintervals.

{\sc Viola and Knill} \cite{VK:02} showed that decoupling can be
achieved by sequentially implementing generators so that they follow a
Eulerian cycle in $\Gamma(G,S)$. An Eulerian cycle is defined as a
cycle that uses each edge exactly once. Because a Cayley graph is
regular, it always has an Eulerian cycle, whose length is necessarily
$N=|G|\, |S|$ (see e.g.\ \cite{Bollobas:98,GR:01} for the definition
of these notions). For our purposes, we use a slightly more general
definition: an Eulerian cycle with multiplicity $\lambda$ is a cycle
that uses each edge exactly $\lambda$ times. Clearly, such an Eulerian
cycle has necessarily length $N=\lambda\, |G|\, |S|$. We will choose
an Eulerian cycle to begin at the identity element of $G$. Therefore,
an Eulerian cycle can be described as a list $(s_1,\ldots,s_N)$ with
entries from $S$. Each entry identifies the edge via which we leave
the vertex.

Decoupling according to an Eulerian cycle $\cC:=(s_1,\ldots,s_N)$ is
defined by setting the cycle time $T_c=N\Delta$ and by choosing the
control propagators $U_c(t)$ as follows:
\begin{equation}\label{eq:eulerian}
U_c\big((j-1) \Delta + \delta\big) = u_{s_j}(\delta)\, 
U_c\big((j-1)\Delta\big)\,
\end{equation}
where $\delta\in [0,\Delta)$ and $u_{s_j}(\delta)$ is defined in
(\ref{eq:control1}) and (\ref{eq:control2}). This decoupling
prescription means that during the $j$th subinterval one choses as a
control Hamiltonian the one that realizes the generator $s_j$, i.e.,
the $j$th element of $\cC$.

The effective Hamiltonian $\bar{H}$ under Eulerian decoupling is
obtained by evaluating the time-average in (\ref{eq:approx}) with the
control propagator being given by (\ref{eq:eulerian}). The resulting
$N$ terms can be partitioned in $|S|$ families, each corresponding to
a fixed generator. Because for each $s$ the cycle $\cC$ contains
exactly $\lambda$ $s$-labeled edges ending at any given vertex $g$,
each family leads to a sum over the group elements as in
(\ref{eq:groupAverage}).

For these reasons the quantum operation $\cQ_\cC$ defined by $\cC$ can
decomposed as
\begin{equation}
\cQ_\cC(X)=\Pi_G(F_S(X))
\end{equation}
with the map $F_S$ implementing an average over both the group
generators and control sub-interval:
\begin{equation}
F_S(X)=\frac{1}{|S|} \sum_{s\in S} \frac{1}{\Delta} \int_0^\Delta
u_s(\tau)^\dagger(s) X u_s(\tau) d\tau\,.
\end{equation}
The link between Eulerian decoupling and bang-bang decoupling by
averaging over $G$ is established in the following theorem. Some
additional compatibility between $\Pi_G$ and $F_S$ is necessary
\cite{VK:02}. Let us repeat all the notions before stating the
theorem. Let $G$ be a group that acts via a faithful, unitary,
projective representation $g\mapsto U_g$ on $\C^d$. The decoupling
group algebra $\cD$ of $G$ is the $\C$-linear span of the matrices
$U_g$.

\begin{theorem}[Eulerian decoupling]\label{th:eulerianDecoupling}${}$\\
Let $X$ be any operator acting on $\C^d$. If the control Hamiltonians
$h_s(t)$ are in the decoupling group algebra, i.e.,
$h_s(\delta)\in\cD$ for all times $\delta\in [0,\Delta]$ and all $s\in
S$, then Eulerian decoupling according an Eulerian cycle $\cC$ as
specified by the the rule in (\ref{eq:eulerian}) has the same effect
as averaging over $G$ as in (\ref{eq:groupAverage}), i.e,
\[
\cQ_\cC(X)=\Pi_G(X)\,.
\]
\end{theorem}
For the proof we refer the reader to \cite{VK:02}. Note that the
bang-bang limit is formally recovered by substituting the map $F_S$ by
the identity map. In the Eulerian approach, at the expense of
lengthening the control cycle, the same $G$-symmetrization can be
attained using only bounded-strength controls. The maximum strengths
achievable in implementing the generators directly bounds the minimum
attainable $T_c$, and therefore the accuracy of the first-order
approximation.

%%%%%%%%%%%%%%%%%%%%%%%%%%%%%%%%%%%%%%%%%%%%%%%%%

\section{Efficient decoupling systems}
In this section we consider a target systems that is composed of $n$
coupled qudits, i.e., its Hilbert space $\cH_S$ is given by the tensor
product $\cH_S= {(\C^d)}^{\otimes n}$. We say that a family of
decoupling schemes is efficient if the number of control operations
grows polynomially with the number of qudits. So far there were only
efficient decoupling schemes using bang-bang controls (see e.g. the
references given in the introduction). The schemes rely on the special
structure of the system Hamiltonians and the couplings to the
environment. It is assumed that the system Hamiltonian is a so-called
few body Hamiltonian. To define this precisely we need to introduce
some notions. For any operator $A$ acting on $\C^d$ we denote by
$A^{(k)}$ the operator that acts as $A$ on the $k$th qudit, i.e.,
$A^{(k)}= \onemat\otimes\cdots\otimes\onemat\otimes A
\otimes\onemat\otimes\cdots\otimes\onemat$.

Let ${\cal B}:=\{\sigma_\alpha\mid \alpha=1,\ldots,d^2\}$ be an basis
of for the vector space $\C^{d\times d}$ of matrices acting on $\C^d$.
We say that an operator $X$ acts on the qudits $k_1,\ldots,k_t$ with
$1\le k_1 < \ldots < k_t \le n$ if it can be expressed as follows
\[
X=\sum_{\alpha_1,\ldots,\alpha_t} 
x_{\alpha_1,\ldots \alpha_t} 
\sigma_\alpha^{(k_1)}\cdots\sigma_\beta^{(k_t)}\,,
\]
for some $x_{\alpha_1,\ldots,\alpha_t}\in\C$.

We assume that the system Hamiltonian is a $t$-body Hamiltonian,
i.e., it can be decomposed as
\begin{equation}\label{eq:pairinteractions}
H_S := \sum_{k_1,\ldots,k_t} H_{k_1,\ldots,k_t}\,,
\end{equation}
where $H_{k_1,\ldots,k_t}$ are traceless operators acting on qudits
$k_1,\ldots,k_t$ only. For $t=2$ one also says that $H$ is a
pair-interaction Hamiltonian. Furthermore, we assume that the environment
couples independently to $t$-tuples of qudits, i.e., we have
\begin{equation}\label{eq:noisegenerators}
H_{SE} = \sum_{k_1,\ldots,k_t} S_{k_1,\ldots,k_t} \otimes 
E_{k_1,\ldots,k_t}
\end{equation}
where $S_{k_1,\ldots,k_t}$ are traceless operators acting on qubits
$k_1,\ldots,k_t$ only and $E_{k_1,\ldots,k_t}$ act on the Hilbert
space $\cH_E$ of the environment.

It will be convenient to use the following definition. Let $X$ be an
arbitrary operator acting on $(\C^d)^{\otimes t}$. We define its
embedding $X^{(k_1,\ldots,k_t)}$ into $(\C^d)^{\otimes n}$ to be the 
operator
\begin{equation}
X^{(k_1,\ldots,k_t)}=
\sum_{\alpha_1,\ldots,\alpha_t} x_{\alpha_1,\ldots \alpha_t} 
\sigma_{\alpha_1}^{(k_1)}\cdots\sigma_{\alpha_t}^{(k_t)}\,,
\end{equation}
where $X=\sum_{\alpha_1,\ldots,\alpha_t} x_{\alpha_1,\ldots \alpha_t}
\sigma_{\alpha_1}\otimes\cdots\otimes\sigma_{\alpha_t}$ is the
expansion of $X$ in the product basis ${\cal B}^{\otimes t}$.

\subsection{Decoupling with bang-bang controls based on orthogonal arrays}

We assume that we can perform bang-bang controls on each qudit
individually. Formally, all control operations are elements of some
finite subset of the group $\cU(d)^{\otimes n}$, where $\cU(d)$
denotes the group of unitary matrices acting on $\C^d$. In the
following we recall how orthogonal arrays may be used to construct
efficient decoupling schemes. Orthogonal arrays appeared first in
statistics where they were used in the design of experiments for
collecting statistical data systematically. We refer the reader to the
books \cite{BJL:99I,CRC,HSS:99} for applications and constructions of
orthogonal arrays. Stollsteimer and Mahler first used orthogonal
arrays (or OAs for short) for the construction of decoupling schemes
and selective coupling schemes \cite{stoll} for qubit systems with
pair-interactions. This method was generalized to qudit systems with
$t$-local interactions in \cite{finite,RW:04}.

\begin{definition}[Orthogonal array of strength $t$]
Let ${\cal A}$ be a finite alphabet and let $n, N \in \N$. An $n\times
N$ array $M$ with entries from ${\cal A}$ is an orthogonal array with
$|{\mathcal A}|$ levels, strength $t$, and multiplicity $\lambda$ if
and only if every $t \times N$ sub-array of $M$ contains each possible
$t$-tuple of elements in ${\cal A}^t$ precisely $\lambda$ times as a
column. We use the notation $OA_\lambda(N,n,s,t)$ to denote a
corresponding orthogonal array. If $\lambda$, $s$, and $t$ are
understood we also use the shorthand notation $OA(N,n)$.
\end{definition}
An important special case arises if the strength $t$ is two. This
means that each pair of elements of ${\cal A}$ occurs $\lambda$ times
in the list $((a_{kj},a_{lj})\mid j=1,\ldots N)$ for $1\le k<l\le n$.
Most of the known construction actually yield arrays of strength two
\cite{HSS:99}. For many physical systems it will be sufficient to
study arrays of small strength since the strength relates to the
degree of the interactions, i.\,e., for pair-interaction Hamiltonians
it is sufficient to consider arrays of strength $t=2$. For an example of 
such orthogonal arrays see \cite{Roetteler:04,RW:04}.

The basic idea is to use an orthogonal array $M$ with parameters
$OA(N,n,d^2,2)$ over an alphabet ${\cal A}$ of size $d^2$. Here $d$
denotes the dimension of the qudits. The elements of ${\cal A}$ are
identified with the elements of the group $G:=Z_d\times Z_d$ that acts
irreducibly on $\C^d$ via the map in (\ref{eq:heisenberg}). The
columns $(g_{1j},\ldots,g_{nj})^T$ of $M$ specify the control
propagators $U_c(t)$ over each of the $N$ equally long subintervals. A
control cycle is defined by
\begin{equation}\label{eq:OAbang}
U_c\big((j-1)\Delta+\tau\big)=
U_{g_{1j}}\otimes U_{g_{2j}}\otimes\cdots\otimes U_{g_{nj}}\,, 
\end{equation}
where $\tau\in [0,\Delta)$, $T_c=N\Delta$ for some $\Delta>0$, and
$j=1,\ldots,N$.

The following theorem shows that the prescription in (\ref{eq:OAbang})
allows to decouple few body Hamiltonians and couplings with the
environment. Let $G$ be an arbitrary finite group. We denote by
$G^{\times t}$ the direct product $G\times\cdots\times G$ having $t$
components.

\begin{theorem}[Decoupling with OAs]${}$\\
Let $G:=Z_d\times Z_d$ and $g\mapsto U_g$ be the irreducible, unitary,
projective representation as in (\ref{eq:heisenberg}). Let
$M=(g_{kj})$ be an $OA(n,N)$ be an orthogonal array of strength $t$
over the group $G$. Let $\Pi_M$ denote the control action that
corresponds to (\ref{eq:OAbang}). Then we have
\begin{equation}
\Pi_M(X^{(k_1,\ldots,k_t)})=\zeromat
\end{equation} 
for an arbitrary traceless operator acting on $(\C^d)^{\otimes t}$ and
any $t$-tuple $(k_1,\ldots,k_t)$ with $1\le k_1<\ldots<k_t\le n$.
\end{theorem}
\proof{The idea is to reduce the problem for each $t$-tuple
$(k_1,\ldots,k_t)$ to the case in (\ref{eq:groupAverage}) by using the
special structure of the operator $X^{(k_1,\ldots,k_t)}$. We have
\begin{widetext}
\begin{eqnarray}
\Pi_M(X^{(k_1,\ldots,k_t)}) 
& = &
\frac{1}{N}
\sum_{j=1}^N 
\big(U_{g_{1j}}\otimes U_{g_{2j}}\otimes\cdots\otimes U_{g_{nj}}\big)^\dagger
\,X^{(k_1,\ldots,k_t)}\,
\big(U_{g_{1j}}\otimes U_{g_{2j}}\otimes\cdots\otimes U_{g_{nj}}\big)
\nonumber\\
& = & 
\left[
\frac{1}{N}
\sum_{j=1}^N 
\big(
U_{g_{k_1,j}}\otimes U_{g_{k_2,j}} \otimes \cdots \otimes U_{g_{k_n,j}}
\big)^\dagger
\, X \,
\big(
U_{g_{k_1,j}}\otimes U_{g_{k_2,j}} \otimes \cdots \otimes U_{g_{k_n,j}}
\big)
\right]^{(k_1,\ldots,k_t)} 
\nonumber\\
& = &
\left[
\frac{\lambda}{N} \sum_{(h_1,\ldots,h_t)\in G^{\times t}}
\big( U_{h_1}\otimes U_{h_2} \otimes \cdots \otimes U_{h_t}\big)^\dagger
\, X \,
\big( U_{h_1}\otimes U_{h_2} \otimes \cdots \otimes U_{h_t}\big)
\right]^{(k_1,\ldots,k_t)} 
\nonumber\\
& = &
\left[
\Pi_{G^{\times t}}(X)
\right]^{(k_1,\ldots,k_t)}  = 
\zeromat^{(k_1,\ldots,k_t)} = \zeromat\,.\label{eq:backToAverage}
\end{eqnarray}
\end{widetext}
The first equality is because $X^{(k_1,\ldots,k_t)}$ acts on the
qudits $k_1,\ldots,k_t$ only. Note that the representation of
$G^{\times t}$ to $\cU(d)^{\otimes t}$ given by
$(h_1,\ldots,h_t)\mapsto U_{h_1}\otimes\cdots\otimes U_{h_t}$ is
irreducible. Since $M$ is an orthogonal array of strength $t$ the list
$((g_{k_1,j},\ldots,g_{k_t,j}))_{j=1}^N$ contains every element of
$G^{\times t}$ exactly $\lambda$ times. Therefore, we average over the
group $G^{\times t}$ acting via an irreducible representation as in
$(\ref{eq:groupAverage})$. This proves eq.~(\ref{eq:backToAverage}).}

It is clear that we can switch off all Hamiltonians of the form in
(\ref{eq:pairinteractions}) and all couplings with an environment of
the form in (\ref{eq:noisegenerators}) with control actions specified
by the above theorem. This is because all operators
$H_{k_1,\ldots,k_t}$ and $S_{k_1,\ldots,k_t}$ have the form
$X^{(k_1,\ldots,k_t)}$ for some traceless operator $X$ acting on
$(\C^d)^{\otimes t}$.

\subsection{Decoupling with bounded strength controls based on Eulerian orthogonal arrays}
Finally, we show how to combine the ideas of Eulerian decoupling and
orthogonal arrays. This is done by introducing the concept of Eulerian
orthogonal arrays.

\begin{definition}[Eulerian orthogonal array]${}$\\
A $n\times N$-matrix $M=(g_{kj})$ with entries from the group $G$ is
said to be an Eulerian orthogonal array of strength $t$ iff for all
$t$-tuples $(k_1,k_2,\ldots,k_t)$ with $1\le k_1<\ldots<k_t\le n$
there is a generating set $S_{k_1,k_2,\ldots,k_t}$ of $G^{\times t}$
such that the list of group elements
\begin{equation}
\big(
(g_{k_1,j},g_{k_2,j},\ldots,g_{k_s,j}\big)_{j=1}^N
\end{equation}
defines an Eulerian cycle in the Cayley graph $\Gamma(G^{\times
t},S_{k_1,k_2,\ldots,k_t})$.
\end{definition}

Note that the above conditions automatically implies that $M$ is a
(usual) orthogonal array of strength $t$.

We assume that we have the ability to implement the group elements
$g\in G$, i.e., to implement the unitaries $U_g$ on the individual
qudits by the application of control Hamiltonians $h_g(t)$ over
$\Delta$ as in (\ref{eq:control1}) and (\ref{eq:control2}). This means
that we have the ability to switch on the control Hamiltonians
$h_g(t)$ on any qudit, i.e.,
$\onemat\otimes\cdots\otimes\onemat\otimes h_g(t) \otimes
\onemat\otimes\cdots\otimes\onemat$.

We define decoupling according to an Eulerian orthogonal array
$M=(g_{kj})$ by setting the cycle time $T_c=N\Delta$ and by assigning
the control propagators as follows:
\begin{eqnarray}\label{eq:eulerOAcycle}
&&
U_c\big(
(j-1)\Delta + \delta
\big) \nonumber\\
& = &
\big(u_{s_{1j}}(\delta) \otimes \cdots \otimes u_{s_{nj}}(\delta)\big)
\,
U_c\big( (j-1)\Delta\big)
\end{eqnarray}
where $\delta\in[0,\Delta)$ and $s_{kj}=g_{kj}^{-1} g_{k,{j+1}}$ for
$j=1,\ldots,N-1$ and $s_{kN}=g_{kN}^{-1} g_{k1}$. The tuples
$(s_{k_1,j},\ldots,s_{k_t,j})$ are the edges in the Eulerian cycle
defined by the rows $k_1,\ldots,k_t$ of $M$.

\begin{theorem}[Decoupling with Eulerian OAs]${}$\\
Let $G:=Z_d\times Z_d$ and $g\mapsto U_g$ be the irreducible, unitary,
projective representation in (\ref{eq:heisenberg}). Let $M=(g_{kj})$
be an Eulerian orthogonal array over $G$ of size $n\times N$ and
strength $t$. Let $\cQ_M$ denote the control action that results from
the control propagator defined in (\ref{eq:eulerOAcycle}). Then we
have
\begin{equation}
\cQ_M(X^{(k_1,\ldots,k_t)})=\zeromat
\end{equation}
for an arbitrary traceless operator $X$ acting on $(\C^d)^{\otimes n}$
and any $t$-tuple $(k_1,\ldots,k_t)$ with $1\le k_1 < \ldots < k_n\le
n$.
\end{theorem}
\proof{Again the idea is to reduce the problem for each $t$-tuple
$(k_1,\ldots,k_t)$ to the case of Theorem~\ref{th:eulerianDecoupling}
by using the special structure of the operator
$X^{(k_1,\ldots,k_t)}$. Let us denote by $\cC_{k_1,\ldots,k_t}$ the
Eulerian cycle in the Cayley graph $\Gamma(G^{\times
t},S_{k_1,\ldots,k_t})$ that is define by the rows $k_1,\ldots,k_t$ of
$M$. Then we have
\begin{eqnarray}
\cQ_M(X^{(k_1,\ldots,k_t)}) & = & 
\left[
\cQ_{\cC_{k_1,\ldots,k_t}}(X)
\right]^{(k_1,\ldots,k_t)} \label{eq:embeddedOA} \\
& = &
\left[
\Pi_{G^{\times t}}\big(F_{S_{k_1,\ldots,k_t}}(X)\big)
\right]^{(k_1,\ldots,k_t)} \nonumber \\
& = &
\zeromat^{(k_1,\dots,k_t)} = \zeromat\,. \nonumber
\end{eqnarray}
Eq.\ (\ref{eq:embeddedOA}) is due to the fact that
$X^{(k_1,\ldots,k_t)}$ acts on qudits $k_1,\ldots,k_t$ only. The
remaining equalities follow from Theorem~\ref{th:eulerianDecoupling}
because all its conditions are satisfied.}

Again it is clear that we can switch off all Hamiltonians of the form
in (\ref{eq:pairinteractions}) and all couplings with an environment
of the form in (\ref{eq:noisegenerators}) with control actions
specified by the above theorem.

\section{Eulerian OAs from linear error correcting codes}
We will construct Eulerian orthogonal arrays using linear error
correcting codes. Let us briefly repeat some basis facts about linear
error correcting codes and their relationship to orthogonal arrays.

A linear code over the finite field $\F_q$ is a $k$-dimensional
subspace of the vector space $\F_q^n$. We consider finite fields of
$q=d^2$ only. In this case the additive group of the finite field
$\F_q$ is isomorphic to $Z_d\times Z_d$; we will again use the
irreducible representation in (\ref{eq:heisenberg}).  The space
$\F_q^n$ is endowed with a metric called Hamming distance. It is
defined as follows: for $x = (x_1, \ldots, x_n) \in \F_q^n$ we have
that ${\rm wt}(x) := |\{ i\in \{1, \ldots, n\} : x_i \not=0\}|$. The
minimum distance of a linear code $C$ is defined by $d = d_{\rm min}
:= \min\{{\rm wt}(c) : c \in C, c\neq {\bf o}\}$, where ${\bf o}$
denotes the zero vector. In this situation we say shortly that $C$ is
an $[n, k, d]_q$ code. We need the fact that a $[n,k]_q$ linear code
can be described by a generator matrix $G$ of size $n\times k$ with
entries from $\F_q$. The matrix $G$ defines the embedding from
$\F_q^k$ to $\F_q^n$; the code words $c\in C$ are the images of the
vectors $m\in\F_q^k$, i.e., $c=G m$. We need one more definition which
is the dual code $C^\perp$ of $C$ defined by $C^\perp := \{ x \in
\F_q^n : x \cdot y = 0 \mbox{ for all } y \in C\}$; the dot product
$x\cdot y$ is given by $\sum_{i=1}^n x_i y_i$. In the following we
refer to the minimum distance $d^\perp$ of the dual code as the dual
distance.

The following theorem \cite[Theorem 4.6]{HSS:99} establishes a close
relationship between orthogonal arrays and error-correcting
codes. This theorem was also used in \cite{Roetteler:04,RW:04}.

\begin{theorem}[OAs from linear codes]\label{th:OAcodes}${}$\\
Let $C$ be a linear $[n, k, d]_q$ code over $\F_q$ with dual distance
$d^\perp$. Arrange the codewords of $C$ into the columns of a matrix
$A \in \F_q^{n\times q^k}$. Then $A$ is an $OA(q^k, n, q, d^\perp-1)$.
\end{theorem}
Now we show how to modify the above construction in order to obtain
Eulerian orthogonal arrays.

\begin{theorem}[Eulerian OAs from linear codes]\label{th:eulerianOA}
Let $C$ be a $[n,k]_q$-code with dual distance $d^\perp$ and $G$ be a
generator matrix for $C$. Let $\cC:=(m_0,\ldots,m_{N-1})$ be an
Eulerian cycle in the Cayley graph $\Gamma(V,S)$ with multiplicity
$1$, where the group is $V:=\F_q^k$ and the generating set is the
group itself, i.\,e., $S:=\F_q^k$. The length of such an Eulerian
cycle is necessarily $N=q^{2k}$. Set $t:=d^\perp-1$. Then the $n\times
N$ matrix $M$ whose columns are defined to be $G m_j$ for
$j=0,\ldots,N-1$ is an Eulerian orthogonal array over $\F_q$ of
strength $t$. Furthermore, we have $S_{k_1,\ldots,k_t}=G^{\times t}$
for all $t$-tuples $(k_1,\ldots,k_t)$ with $1\le k_1<\ldots<k_t\le n$.
\end{theorem}
\proof{Since $\cC=(m_0,\ldots,m_{N-1})$ is an Eulerian cycle all
elements of $V=\F_q^k$ appear exactly $q^k$ (corresponding to the size
$|S|=q^k$ of the generating set $S$) times in $\cC$. Therefore, the
column vector $Gm$ appears exactly $q^k$ times in $M$ for all
$m\in\F_q^k$. It now follows from Theorem~\ref{th:OAcodes} that $M$ is
an orthogonal array; its multiplicity is just $q^k$ times the
multiplicity of an OA constructed based on Theorem~\ref{th:OAcodes}.

Let $m$ be an arbitrary element of $V=\F_q^k$ and $I_m:=\{j \mid m_j
=m\}$. Then every element of $S=\F_q^k$ appears exactly once in the
list $(m_{j+1} - m_j \mid j\in I_m)$ because $\cC$ is an Eulerian
cycle in $\Gamma(V,S)$ with multiplicity one (the addition is done
modulo $N$). Consequently, the list of transitions that occur in $M$
from all columns of the form $G m$, i.e., $(G m_{j+1} - G {m_j} \mid
j\in I_m)$ is independent of $m$ and is equal (up to reordering the
columns) to the orthogonal array $M':=(G e\mid e\in \F_q^k)$; it
follows from Theorem~\ref{th:OAcodes} that $M'$ is an orthogonal
array. This proves that $M$ is Eulerian and also that
$S_{k_1,\ldots,k_t}=G^{\times t}$ for all $t$-tuples since $M'$ is an
orthogonal array of strength $t$.}

Note that our construction is closely related to {\sc R{\"o}tteler}'s
construction \cite{Roetteler:04}. In that paper Hamiltonian cycles in
the Cayley graph $\Gamma(\F_q^k,S)$ are used, where the generating set
is $S$ given by the $k$ coordinate vectors. The motivation behind this
construction is to reduce the number of different control pulses in
bang-bang decoupling.

Let us now explain how to construct decoupling schemes for general
$t$-body Hamiltonians acting on $n$ qudits with bounded controls based
on Theorem~\ref{th:eulerianOA}. To obtain a decoupling scheme using a
minimal number of pulses we have to find a code $[n,k]_q$ such that
$k$ is minimal and the dual distance $d^\perp$ is at least $t+1$. This
may be formulated in terms of the dual code which has parameters
$C^\perp=[n,n-k,d^\perp]_q$. The dual code $C^\perp$ should contain
the maximally possible number of code words for given $n$ and
$d^\perp$. This question is one of the central optimization problems
in the theory of error correcting codes. To find such optimal or best
known codes one could e.g.\ use the computer algebra system {\rm
MAGMA} \cite{magma} that contains a table of best known linear codes
(i.e., with the maximal number of code words) for given length and
minimal distance.

We now consider a quantum system consisting of $n$ qubits which are
governed by a pair-interaction Hamiltonian. For such a system we can
construct decoupling schemes using $N$ pulses from an orthogonal array
$OA(N, n, $4$, 2)$. Hence, in order to apply Theorem~\ref{th:OAcodes}
and \ref{th:eulerianOA} we have to find a code $C$ of linear codes
over $\F_4$ for which the parameters are $[n, k, d]$ and for which the
dual distance is at least $3$. This can be done with the help of
Hamming codes \cite{Roetteler:04,RW:04}. For every $m\in \N$ there is
an orthogonal array $OA(4^m,(4^m-1)/3,4,2)$. The columns of this OA
are codewords of the dual code of a Hamming code. The corresponding
Eulerian orthogonal array has parameters $OA(16^m,(4^m-1)/3,4,2)$.

To obtain a decoupling scheme for a quantum system consisting of $n$
qubits, where $n$ is an arbitrary natural number, i.\,e., not
necessarily of the form $n=(4^m-1)/3$ we proceed as follows: first let
$m\in \N$ be the unique integer such that $n \leq \frac{4^m-1}{3} \leq
4n$. Then construct the orthogonal array with parameters $OA(4^m,
(4^m-1)/3,4,2)$ for bang-bang controls and the Eulerian orthogonal
array with parameters $OA(16^m,(4^m-1)/3,4,2)$ for bounded-strength
controls, respectively. These results shows that the complexity of
decoupling for general pair-interactions Hamiltonians acting on $n$
qubits scales at most linearly in $n$ for bang-bang controls and at
most quadratically in $n$ for bounded-strength controls, respectively.

{\it Conclusions and Discussions}.--- We have shown that it is
possible to construct decoupling schemes using bounded-strength
controls for composite multipartite qudit systems with the help of
Eulerian orthogonal arrays. Our concept of Eulerian orthogonal arrays
merges the desirable properties of usual orthogonal arrays (that were
use to construct efficient decoupling schemes with bang-bang controls)
and Eulerian cycles that are at the heart of {\sc Viola and Knill}'s
Eulerian decoupling method. We have shown how to construct efficient
Eulerian orthogonal arrays based on linear error correcting codes. It
would be interesting to find new construction of Eulerian orthogonal
arrays that might yield decoupling schemes with a smaller number of
pulses.

The author would like to thank Anja Groch and Markus Grassl for
helpful discussions. This work has been supported by the National
Science Foundation under grant EIA-$0086038$ through the Institute for
Quantum Information at the California Institute of Technology and
the BMBF-project 01/BB01B.

\newcommand{\etalchar}[1]{$^{#1}$}

\end{document}